\pdfoutput=1
\documentclass[11pt]{article}

\usepackage[final]{acl}

\usepackage{times}
\usepackage{latexsym}
\usepackage{xcolor}
\usepackage{booktabs}
\usepackage{longtable}
\usepackage{hyperref}

\usepackage[T1]{fontenc}
\usepackage[utf8]{inputenc}

\usepackage{microtype}

\usepackage{inconsolata}

\usepackage{listings}
\usepackage{booktabs}
\usepackage{multirow}
\usepackage{listings} % Load the listings package
\usepackage{graphicx}
\usepackage{array}
\usepackage{lipsum} % For dummy text
\usepackage{svg}
\usepackage{float} % For H placement
\usepackage{placeins} % For float barriers
\usepackage{algorithm}
\usepackage{algorithmic}
\usepackage{caption}
\usepackage{subfigure}
\usepackage{subcaption}
\usepackage{tabularx}
\usepackage{url}
\usepackage{amsmath}
\usepackage[para,online,flushleft]{threeparttable}
\title{A Simple but Effective Elaborative Query Reformulation Approach for Natural Language Recommendation}

\author{Qianfeng Wen\thanks{Equal contribution}\textsuperscript{\textnormal{1}},
  Yifan Liu\footnotemark[1]\textsuperscript{\textnormal{1}}, Justin Cui\footnotemark[1]\textsuperscript{\textnormal{1}},\\ 
  \textbf{Joshua Zhang\textsuperscript{\textnormal{1}},
  Anton Korikov\textsuperscript{\textnormal{1}},
  George-Kirollos	Saad\textsuperscript{\textnormal{1}},
  Scott Sanner\textsuperscript{\textnormal{1}}}\\
  \textsuperscript{1}University of Toronto, Canada\\
}

\begin{document}
\maketitle
\begin{abstract}
Natural Language (NL) recommender systems aim to retrieve relevant items from free-form user queries and item descriptions. Existing systems often rely on dense retrieval (DR), which struggles to interpret challenging queries that express broad (e.g., “cities for youth-friendly activities”) or indirect (e.g., “cities for a high school graduation trip”) user intents. While query reformulation (QR) has been widely adopted to improve such systems, existing QR methods tend to focus \textit{only} on expanding the range of query subtopics (breadth) or elaborating on the potential meaning of a query (depth), but not both. In this paper, we propose \textbf{EQR} (Elaborative Subtopic Query Reformulation), a large language model-based QR method that combines \textit{both} breadth and depth by generating potential query subtopics with information-rich elaborations. We also introduce three new natural language recommendation benchmarks in travel, hotel, and restaurant domains to establish evaluation of NL recommendation with challenging queries. Experiments show EQR substantially outperforms state-of-the-art QR methods in various evaluation metrics, highlighting that a simple yet effective QR approach can significantly improve NL recommender systems for queries with broad and indirect user intents.
\end{abstract}
\section{Introduction}
\label{sec:intro}

% \begin{figure*}[t!]
%     \centering   
%     \includegraphics[width=1.0\textwidth]{resources/merge.pdf}
%     \caption{Examples of broad (\textit{Cities for youth-friendly activities}, right) and indirect (\textit{Cities for a high school graduation trip}, left) queries. Comparisons are made across different QR methods discussed in \autoref{qr}. The top 5 recommendations are shown, with irrelevant results marked in red with an {\color{red} $X$}. \textbf{Q2E} captures subtopic breadth but lacks depth, \textbf{Query2Doc} provides depth without breadth, and \textbf{EQR} achieves both.}
%     \label{fig:examples}
% \end{figure*}

Natural Language (NL) Recommender Systems~\cite{kang2017nlrecsys} aim to generate item recommendations from user-issued NL queries. These systems assume that the query itself encodes user preferences and provides the signals necessary for personalization, which traditional recommenders typically infer from interaction history~\cite{afzali2023pointrec}. Each item is typically associated with multiple descriptive passages (e.g., reviews, wiki pages), and effective NL recommendation requires reasoning over multiple textual sources that capture different aspects of an item~\cite{kemper2024retrieval, wen2024elaborative}.

\begin{figure}[t!]
    \centering
    \includegraphics[width=1.0\linewidth]{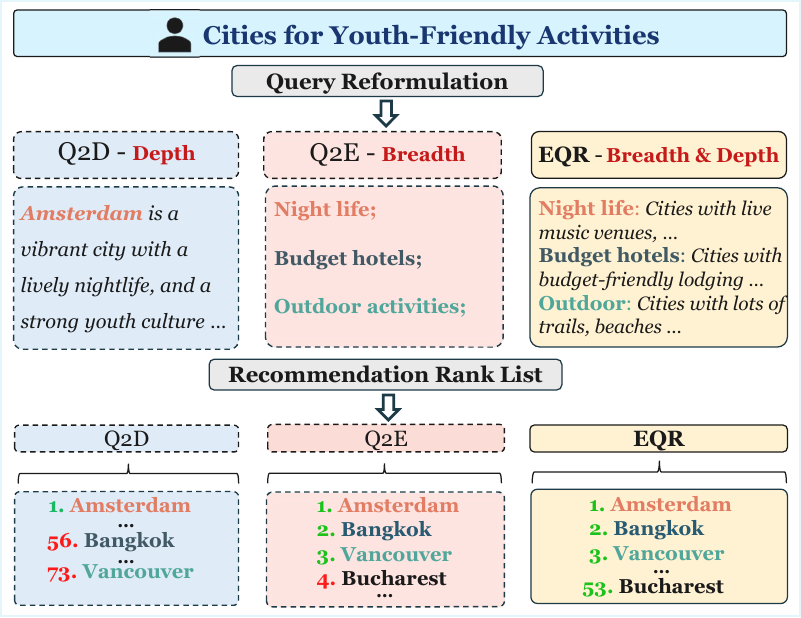}
    \caption{
    Example recommendation results for the query \textit{"Cities for youth-friendly activities"} under different QR methods. We show results for four representative cities: Amsterdam (known for nightlife), Bangkok (known for vibrant street life and budget accommodations), and Vancouver (known for outdoor activities) are part of the ground truth, while Bucharest—although budget-friendly—is not considered youth-friendly. \textbf{Q2D} focuses solely on \textit{depth}, generating an in-depth reformulation that highlights Amsterdam but fails to surface other relevant candidates. \textbf{Q2E} emphasizes \textit{breadth} by listing diverse keywords, but incorrectly ranks Bucharest highly due to its affordability. In contrast, \textbf{EQR} effectively distinguishes ideal and non-ideal candidates by combining both \textit{breadth} and \textit{depth} in its reformulation.
    }
    \label{fig:eqr_hook}
\end{figure}

\begin{figure*}[t]
    \centering
    \includegraphics[width=0.9\textwidth]{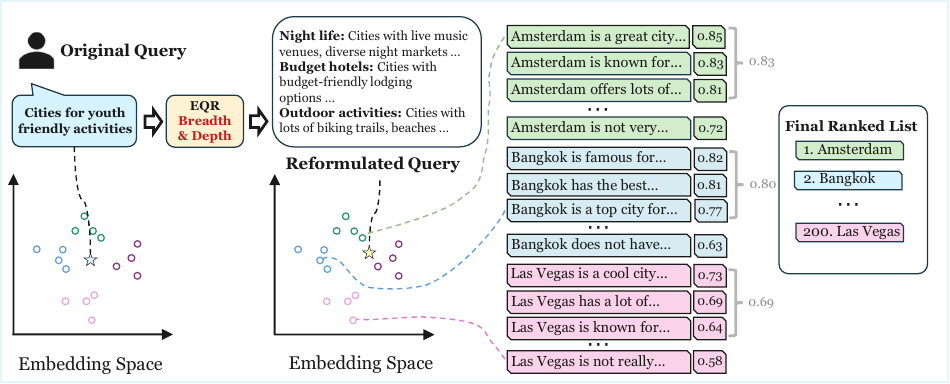}
    \caption{
    Pipeline overview of an NL recommender system with LLM-driven query reformulation (QR). Passage scores represent the cosine similarity between the reformulated query and each passage in the embedding space. Item-level scores are computed by averaging the top-$n$ passage scores.
    }
    \label{fig:eqr_pipeline}
\end{figure*}

However, matching NL queries to multiple textual aspects is challenging for standard dense retrieval (DR) methods, especially for broad queries that imply multiple subtopics (e.g., \textit{``cities for youth-friendly activities''}) and indirect queries that require inference beyond the query text (e.g., \textit{``cities for a high school graduation trip''}), as they lack the reasoning capabilities needed to bridge these implicit user intents to multiple textual aspects without explicit query cues \citep{karpukhin2020dense, liu2025ma}.

To address this, prior work has explored Query Reformulation (QR)~\cite{InferringQueryIntentfromReformulationsandClicks, AutomaticQueryExpansion}, with recent advances leveraging Large Language Models (LLMs)~\cite{languagelearningAreFewShotLearners, 2023GenQR}. LLM-based QR methods typically focus on either: (a) expanding queries by adding diverse keywords to improve subtopic \textit{breadth}~\cite{q2dq2e, genqr}, or (b) generating paraphrases or relevant passages to enhance conceptual \textit{depth}~\cite{hyde, q2dq2e, query2doc, gqrgqe, 2023GenQR}.

We hypothesize that effective NL recommendation requires addressing \textit{both} breadth and depth. Moreover, we observe that LLMs' general reasoning capabilities~\cite{tafjord2020proofwriter, yao2024tree} can support expansions that simultaneously cover a diverse set of subtopics (breadth) and enrich each subtopic with detailed, inferential content (depth), improving retrieval for broad and indirect queries. Our contributions are as follows:

\begin{itemize}
    \item We propose \textbf{EQR} (Elaborative Subtopic Query Reformulation) \footnote{Code available at: \url{https://github.com/cuijustin0617/query_driven_rec_datasets}}, an LLM-based QR method that infers multiple subtopic (\textit{breadth}) and provides information-rich elaborations for each (\textit{depth}). As illustrated in \autoref{fig:eqr_hook}, \textbf{EQR} better combines breadth and depth compared to existing QR methods.
    
    \item We introduce three large-scale, LLM-curated benchmark datasets for NL recommendation spanning the travel destination, hotel, and restaurant domains. Empirical results demonstrate that \textbf{EQR}, based on a simple and intuitive prompting idea, consistently outperforms all baseline QR methods across these datasets.\footnote{Data available at: \url{https://huggingface.co/datasets/cuijustin0617/NLRec}}

\end{itemize}

% We conclude with a comparative evaluation of QR methods on \textbf{TravelDest} and find that \textbf{EQR} outperforms existing methods in terms of recall and precision metrics, thus improving the ability of Travel RSs to address challenging broad and indirect NL queries. 

\section{Related Work}

\subsection{Natural Language Recommender System}

Recent years have seen growing interest in natural language (NL) recommendation, where users issue free-form textual requests to retrieve relevant items. Early studies such as Kang et al.~\cite{kang2017nlrecsys} analyzed how users naturally express recommendation needs, which highlights the potential for query-driven personalization. NL recommendation is closely related to narrative-driven recommendation~\cite{ndr2017}, initially formalized by Bogers and Koolen~\cite{ndr2017} for book recommendation, where users describe preferences through long-form narrative queries. Later work extended NL recommendation to additional domains, including movies~\cite{bogers2018movie}, video games~\cite{bogers2019game}, and points of interest~\cite{afzali2023pointrec}. While early formulations incorporated prior user interactions, more recent approaches such as Afzali et al.~\cite{afzali2023pointrec} showed that rich contextual cues embedded within narrative queries alone can support effective recommendation without relying on historical user data.

\subsection{Query Reformulation}
While query reformulation (QR) has been studied extensively over past decades~\cite{deerwester1990indexing, dumais1988using, 1971PRF, 1990PRF, 2002PRF}, recent advances in large language models (LLMs) have introduced new capabilities for reformulating queries using internalized language knowledge. Modern LLM-based QR methods enable more flexible and semantically rich reformulations compared to traditional expansion techniques. Among them, \emph{keyword-based} and \emph{relevant answer passage-based} approaches have received significant attention. Keyword-based methods expand the coverage of the original query by generating additional relevant terms~\cite{q2dq2e, genqr, rashid2024progressive}, often guided by pseudo-relevance feedback or iterative keyword generation. Relevant answer passage-based methods reformulate queries by retrieving or generating information-dense passages that reflect the potential intent behind the original query~\cite{q2dq2e, query2doc, hyde}, aiming to enrich the semantic depth available to retrieval systems.

However, most existing QR methods focus on either expanding subtopic \emph{breadth} or enhancing conceptual \emph{depth}, but rarely both. This limits their effectiveness for complex NL queries requiring both broad coverage and rich elaboration. Our work addresses this gap by proposing a method that jointly targets breadth and depth in reformulation, improving alignment with multi-aspect item representations.

\section{Methodology}

\subsection{Natural Language Recommender System}  
Let \( q \) be an NL query, and let \(\mathcal{I}\) be the set of all items. Each item \( i \in \mathcal{I} \) is associated with a set of passages \(\mathcal{P}_i = \{ p_1, p_2, \ldots, p_m \} \), where each \( p_j \) is a description or review of item \( i \).  

The goal of a NL recommender system is to produce a ranked list \(\mathcal{S}\) of items \( i \in \mathcal{I} \) based on their relevance to the query. A simple yet effective scoring procedure is defined as follows:

\begin{algorithm}[t!]
\caption{Item Scoring Algorithm}
\label{alg:dest_score}
\begin{algorithmic}[1]
\STATE $q' \gets \text{Reformulate}(q)$ 
\COMMENT{See \autoref{qr}}
\FOR{each item $i \in \mathcal{I}$}
    \STATE $\mathbf{q'} \gets \text{Encode}(q')$
    \FOR{each passage $p_j \in \mathcal{P}_i$}
        \STATE $\mathbf{p}_j \gets \text{Encode}(p_j)$
        \STATE $\text{score}(q', p_j) \gets \text{cos}(q', p_j)$ \COMMENT{dense similarity} 
        % or \text{BM25}$(q',\!c_j)$ \COMMENT{sparse}
    \ENDFOR
    \STATE $\mathcal{P}_{q'} \gets$ top-$n$ passages $\{ p_1, p_2, \ldots, p_n \}$ by $\text{score}(q', p_j)$ 
    \STATE $\text{score}(i) \gets \frac{1}{n} \sum_{p_j \in \mathcal{P}_{q'}} \text{score}(q', p_j)$ \COMMENT{Average of top-$n$}
\ENDFOR
\STATE $\mathcal{S} \gets$ Sort items $i$ by $\text{score}(i)$ in descending order
\end{algorithmic}
\end{algorithm}

\subsection{Query Reformulation} \label{qr}
\begin{figure}[t!]
    \centering
    \includegraphics[width=0.8\linewidth]
    {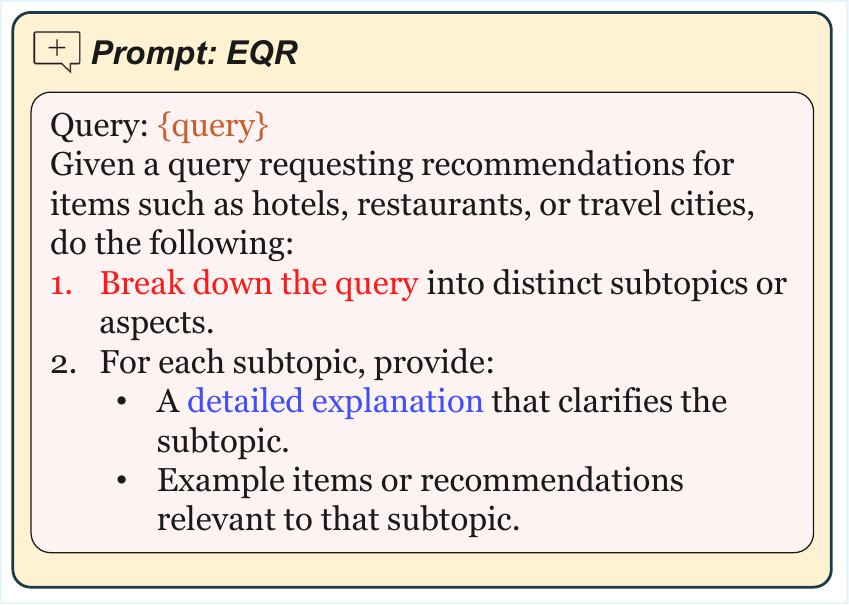}
    \caption{Prompts for EQR discussed in \autoref{qr}. The first bullet point (in {\color{red} red}) adds breadth to the query, while the second bullet point (in {\color{blue} blue}) introduces depth.}
    \label{fig:prompt}
\end{figure}

In this work, we fix the structure of the Query-driven Recommender as in Algorithm \autoref{alg:dest_score} while experimenting with the impact of different QR methods to implement Line 1, defined as follows:
%and comparatively evaluate the efficacy of different QR methods to instantiate Line 1:
%We introduce the following reformulation function setups:

\begin{description}
\item[No QR]\!\!: $q' = q$, which means no QR is applied.

\item[Q2E]\!\cite{q2dq2e}: $q' = q + \text{LLM}(q, \text{Q2E-prompt})$, which expands the original query by adding multiple keywords using the LLM.

\item[Query2Doc]\!\cite{query2doc}: $q' = q + \text{LLM}(q, \text{Query2Doc-prompt})$, which generates relevant answer passages from the query using the LLM and concatenates them with the original query.

% \item[GQR]\!\cite{gqrgqe}: $q' = q + \text{LLM}(q, \text{GenQR-prompt})$, which paraphrases the original query using the LLM.

\item[EQR]\!\!: $q' = q + \text{LLM}(q, \text{EQR-prompt})$, which generates $k$ subtopic elaboration paragraphs from the query using the LLM. See \autoref{fig:prompt} for a detailed prompt and \autoref{eqr} for a detailed discussion on \textbf{EQR}.
\end{description}

% \begin{figure}[t!]
%     \centering
%     \includegraphics[width=1.0\linewidth]{resources/prompt.pdf}
%     \caption{LLM prompts for various QR methods discussed in Section \autoref{qr}, with LLM output shown in Figure \autoref{fig:examples} using two broad and indirect query examples. Q2E \cite{q2dq2e} (top-left), Query2Doc \cite{query2doc} (top-right), GenQR \cite{gqrgqe} (bottom-left), and EQR (bottom-right).}
%     \label{fig:prompt}
% \end{figure}

\subsection{EQR: Elaborative Subtopic Query Reformulation} \label{eqr}

The general idea behind \textbf{EQR} as motivated in \autoref{sec:intro} is to infer multiple subtopics from an original query (i.e., breadth) while elaborating each with information-rich content using the LLM's general reasoning abilities (i.e., depth). 
%This novel method combines existing breadth-focused keyword-based methods and depth-focused paraphrase-based methods to better capture user intent behind the challenging, broad and indirect queries.

Specifically, \textbf{EQR} involves two steps designed to address both breadth and depth, as detailed below:

\begin{description}
    \item[{\color{red} \textit{[Breadth]}} Number of Subtopics:] It first generates a set of distinct subtopics from a given NL query $q$, which adds a breadth aspect to capture a wider range of relevant or latent subtopics compared to answer-based and paraphrase-based methods \cite{hyde, query2doc, q2dq2e, gqrgqe}. 
    
    \item[{\color{blue} \textit{[Depth]}} Elaboration of Subtopics:]     Each subtopic is then expanded into an information-rich description, denoted \( e_1, e_2, \\ \cdots, e_k \). This step provides more detailed, logically entailed connections between the query and inferred subtopics, offering greater depth compared to keyword-based methods \cite{q2dq2e, genqr}.
\end{description}

% \textbf{EQR} employs the LLM prompt shown in \autoref{fig:prompt}. The reformulated query $q'$ is a keyword merge $q' \! = \! \text{concat}(q, e1, e2, \cdots, e_k)$ for {\bf sparse retrieval via BM25} ~\cite{robertson1994bm25}, or a [SEP]-delimited text concatenation $q' \! = \! \text{concat}(q, \text{[SEP]}, e_1, \cdots, \text{[SEP]}, e_k)$ for LLM encoding and {\bf dense retrieval via cosine similarity} (cf. Line 6 in Algorithm \autoref{alg:dest_score}).

The new query $q'$ is constructed by concatenating $q$ with $e_1, \cdots, e_k$, separated by [SEP] tokens, which is a convention in LLM-based QR method for DR \cite{convgqr, query2doc} $q' = \text{concat}(q, \text{[SEP]}, e_1, \ldots, \text{[SEP]}, e_k)$

\section{Benchmark Datasets for NL Recommendation}
\label{sec:datasets}

\begin{table*}[ht]
\centering
\resizebox{\linewidth}{!}{%
\begin{tabular}{lcccc|cccc|cccc}
\toprule
 & \multicolumn{4}{c}{\textbf{TravelDest}}
 & \multicolumn{4}{c}{\textbf{TripAdvisor Hotel}}
 & \multicolumn{4}{c}{\textbf{Yelp Restaurant}} \\
\cmidrule(lr){2-5}\cmidrule(lr){6-9}\cmidrule(lr){10-13}
 & NDCG@10 & NDCG@30 & P@10 & P@30
 & NDCG@10 & NDCG@30 & P@10 & P@30
 & NDCG@10 & NDCG@30 & P@10 & P@30 \\
\midrule
\multicolumn{13}{c}{\textbf{all-MiniLM-L6-v2}}\\
\midrule
\textbf{No QR}       & 0.564 & 0.503 & 0.549 & 0.473
                     & 0.255 & 0.304 & 0.207 & 0.151
                     & 0.474 & 0.435 & 0.455 & 0.385 \\ 
\textbf{Q2E}         & 0.628 & 0.532  & 0.605 & \underline{0.512}
                     & \underline{0.327} & \underline{0.377} & \underline{0.261} & \underline{0.172}
                     & \underline{0.572} & \underline{0.507} & \underline{0.547} & \underline{0.422} \\ 
\textbf{Query2Doc}   & \underline{0.672} & \underline{0.553} & \underline{0.642} & 0.509
                     & 0.288 & 0.344 & 0.218 & 0.166
                     & 0.422 & 0.385 & 0.404 & 0.345 \\ 
\textbf{EQR (Ours)}  & \textbf{0.719} & \textbf{0.618} & \textbf{0.679} & \textbf{0.563}
                     & \textbf{0.371} & \textbf{0.404} & \textbf{0.296} & \textbf{0.185}
                     & \textbf{0.619} & \textbf{0.538} & \textbf{0.584} & \textbf{0.465} \\ 
\midrule
\multicolumn{13}{c}{\textbf{e5-small-v2}}\\
\midrule
\textbf{No QR}       & 0.579 & 0.523 & 0.565 & 0.494
                     & 0.272 & 0.313 & 0.216 & 0.151
                     & 0.580 & 0.503 & 0.546 & 0.428 \\ 
\textbf{Q2E}         & 0.655 & 0.548 & 0.636 & \underline{0.519}
                     & \underline{0.324} & \underline{0.379} & \underline{0.259} & \underline{0.172}
                     & \underline{0.616} & \underline{0.543} & \underline{0.572} & \underline{0.471} \\ 
\textbf{Query2Doc}   & \underline{0.691} & \underline{0.582} & \underline{0.644} & 0.518
                     & 0.284 & 0.342 & 0.225 & 0.166
                     & 0.525 & 0.472 & 0.500 & 0.415 \\ 
\textbf{EQR (Ours)}  & \textbf{0.721} & \textbf{0.613} & \textbf{0.690} & \textbf{0.559}
                     & \textbf{0.358} & \textbf{0.397} & \textbf{0.273} & \textbf{0.180}
                     & \textbf{0.657} & \textbf{0.572} & \textbf{0.618} & \textbf{0.493} \\ 
\bottomrule
\end{tabular}}

% \end{table*}

\caption{Comparative performance of QR methods using different dense retrieval embedding models across the three benchmark datasets. Best results are highlighted in bold, and second-best results are underlined. The results show that \textbf{EQR} consistently outperforms other LLM-based QR methods across datasets and embedding models.}
\end{table*}
\label{tab:experiment_results}

Despite growing interest in NL recommendation \cite{kang2017nlrecsys, ndr2017, bogers2018movie, bogers2019game, afzali2023pointrec}, there is a lack of benchmark datasets that specifically evaluate dense retrieval (DR) under challenging conditions where user intent is implicitly expressed through broad or indirect queries, and items are described through multiple diverse textual sources. This setting presents unique difficulties for matching queries to relevant content, as it requires reasoning across multi-aspect item representations without explicit query cues. 

To address this gap, we release three large-scale benchmark datasets from diverse domains—\texttt{TravelDest}, \texttt{TripAdvisor Hotel}, and \texttt{Yelp Restaurant}—designed to rigorously evaluate NL recommender systems under these challenging conditions. Each dataset includes a set of challenging NL queries for item recommendation, a collection of target items (e.g., travel destinations, hotels, and restaurants), a set of textual passages associated with each item, and ground truth relevance labels for each query. Detailed information for each dataset is as follows:

\begin{itemize}
    \item \textbf{TravelDest} – 100 queries and 775 travel destinations, each associated with a \texttt{WikiVoyage} page~\footnote{Content used under the Creative Commons Attribution-ShareAlike 4.0 International License. See: \url{https://creativecommons.org/licenses/by-sa/4.0/legalcode}}. We treat each section in the page (e.g., \emph{History}, \emph{Attractions}, \emph{Getting Around}) as a separate passage.
    
    \item \textbf{TripAdvisor Hotel} – 100 queries each for Philadelphia and New Orleans, with 1152 hotels in total. Each hotel is associated with a set of review snippets describing amenities, location, service quality, and other relevant attributes, collected from \texttt{TripAdvisor}~\footnote{Data used in accordance with TripAdvisor's content policy. See: \url{https://tripadvisor.mediaroom.com/US-resources}}.
    
    \item \textbf{Yelp Restaurant} – 100 queries for each of New York, Chicago, London, and Montreal, with 589 restaurants in total. Each restaurant is paired with user reviews covering aspects such as menu items, ambiance, and service, sourced from \texttt{Yelp}~.\footnote{Data used in accordance with Yelp's Terms of Service. See: \url{https://terms.yelp.com/tos/en_us/20240710_en_us/}}
\end{itemize}

% \paragraph{Query Generation}
% For each domain we write 100 entity-free, free-form requests that describe scenarios, moods, or abstract attributes rather than naming specific items.  The set mixes:  
% (i) broad goals (e.g., a city “for a high-school graduation trip”);  
% (ii) indirect experiential prompts that hinge on nuance (“Where can I find poetic fusion without the pretension?”); and  
% (iii) blended constraints combining ambience, budget, or context.  
% By omitting explicit keywords and emphasizing experiential intent, these queries require models to reason semantically rather than match surface forms.

\paragraph{Relevance Label}  
We adopt an LLM-based approach to construct ground truth relevance labels for each query. Specifically, for every query-item pair, we use \texttt{Gemini-2.0-flash}~\cite{gemini2024}, a language model distinct from the one used later for LLM-driven query reformulation. The model is prompted with the query, the candidate item, and all passages associated with the item using the \texttt{UMBRELA} prompting framework~\cite{li2024umbrella}, and produces a binary label indicating whether the item is an \emph{ideal candidate} (label = 1) or a \emph{non-ideal candidate} (label = 0) with respect to the information need expressed in the query. All items labeled as ideal candidates are treated as ground truth for that query.\footnote{Code for dataset curation is available at \href{https://github.com/cuijustin0617/query_driven_rec_datasets}{github}.}

To assess the quality of the LLM-generated labels, we select a subset of 42 queries from the \texttt{TravelDest} dataset and recruit domain experts in travel to manually annotate ground truth relevance. Comparing the LLM-generated labels with expert annotations yields a Cohen’s $\kappa$ of 0.39, indicating \emph{fair agreement}.

\begin{table}[ht]
    \centering
    \resizebox{0.48\textwidth}{!}{%
    \begin{tabular}{lcccc}
        \toprule
        \textbf{Dataset} & \textbf{\# Queries} & \textbf{\# Items} & \textbf{\# Passages} & \textbf{\# Labels} \\
        \midrule
        TravelDest & 100 & 775 & 126,400 & 4,887 \\
        TripAdvisor Hotel & 100 & 589 & 133,759 & 2\,356 \\
        Yelp Restaurant & 100 & 1,152 & 283,658 & 11,726 \\
        \bottomrule
    \end{tabular}%
    }
    \caption{Statistics of our benchmark datasets.}
    \label{tab:datasets}
\end{table}

\section{Experiments}
\subsection{Setup} \label{exp}
We evaluate dense retrieval using cosine similarity with two widely used BERT-based sentence encoders: \texttt{E5}~\cite{wang2022text} and \texttt{MiniLM}~\cite{wang2020minilm}, both implemented via the HuggingFace \texttt{sentence-transformers}~\cite{huggingface2024sentencetransformers}. To ensure consistency across methods, all QR variants use \texttt{GPT-4o}~\cite{achiam2023gpt} as the common LLM for query reformulation. We set the number of top-$n$ passages for aggregation to 50. Evaluation is performed using standard metrics, including NDCG and Precision at ranks 10 and 30.

% \subsection{Metrics}
% \begin{description}
% \item[GenQR] \cite{2023GenQR}: Paraphrases the original query using the LLM, a paraphrase-based method.

% \item[Q2E] \cite{q2dq2e}: Expands the original query by adding multiple keywords using the LLM, a keyword-based method.\footnote{Originally designed for sparse retrieval, adapted for dense retrieval as well by concatenating each keyword with [SEP] tokens.}

% \item[Query2Doc] \cite{query2doc}: Generates relevant answer passages from the query using the LLM and concatenates them with the original query, a relevant answer passage-based method.

% \item[EQR]: Generates $k$ subtopic elaboration paragraphs from the query using the LLM, a subtopic elaboration method.\footnote{We experimented with different levels of $k$ and reported the best results. See the appendix for more details.}
% \end{description}

% We primarily focus on Recall metrics to ensure the minimal number of relevant items are missed and report Recall@10 and Recall@50. Additionally, we report R-Precision to verify system precision relative to the actual ground truth size and mAP@10 and mAP@50 to assess system ability to rank more relevant items earlier in the results list.
%ensure that the system retrieves relevant destinations early in the results list.

\subsection{Results} \label{res}

\autoref{tab:experiment_results} presents comparative results for all QR methods across the three benchmark datasets. LLM-based QR methods consistently outperform the baseline dense retrieval (\textbf{No QR}), confirming that LLM-driven reformulation enhances retrieval effectiveness, particularly for broad and indirect queries. 

However, performance varies across datasets due to differences in granularity. \texttt{TravelDest} features destination-level queries, allowing LLMs to leverage internal knowledge and generate effective reformulations. In contrast, \texttt{Yelp Restaurant} and \texttt{TripAdvisor Hotel} focus on finer-grained entities like individual hotels and restaurants, where LLMs have limited knowledge coverage, making detailed reformulations more difficult.

Consistent with these differences, \textbf{Query2Doc} performs best on \texttt{TravelDest} by providing semantically rich, in-depth reformulations, while \textbf{Q2E} performs best on the other two datasets by providing broader keyword-based expansions that align more effectively with queries targeting fine-grained items.

\textbf{EQR} effectively combines the strengths of both approaches and achieves superior performance across all datasets, metrics, and embedding models. These results demonstrate that enhancing both subtopic breadth and semantic depth in query reformulation leads to more robust and generalizable improvements.

\section{Conclusion}
We presented Elaborative Subtopic Query Reformulation (\textbf{EQR}), an LLM-based QR approach that enhances both breadth and depth by generating multiple, information-rich subtopic elaborations for broad or indirect queries. Additionally, we introduced three query-driven recommender system benchmark datasets—spanning travel cities, hotel, and restaurant domains—to facilitate evaluation of query-reformulation methods and promote further research in query-driven recommendation. \textbf{EQR} consistently achieved state-of-the-art performance across various evaluation metrics, datasets, and retriever types.

\bibliography{main}

\newpage
\appendix

\section{Ablation Studies}
\begin{figure*}[!t]
    \centering
    \includegraphics[width=1.0\linewidth]{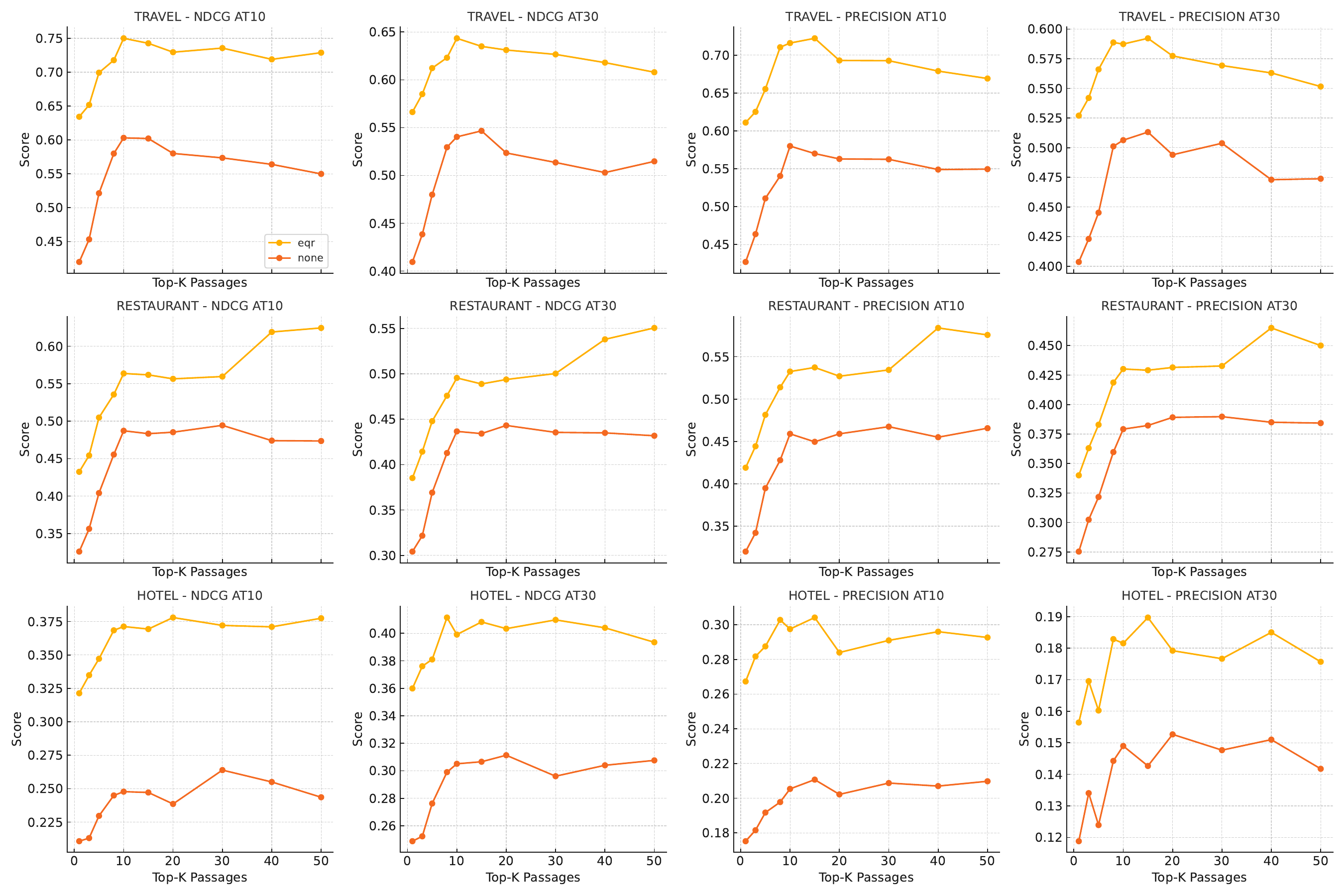}
    \caption{Top-n parameter performance among all datasets}
    \label{fig:performance_metrics}
\end{figure*}

In this section, we examine how varying the top-$n$ parameter affects the performance of \textbf{EQR} (see Fig \ref{fig:performance_metrics}). In the main experiments, we fixed this value at 50; here, we conduct an ablation study to demonstrate that 50 serves as a conservative lower bound. We observe that performance typically improves as $n$ increases up to a point, after which it begins to decline, indicating an optimal range for top-$n$ selection.

\section{Human Label Alignment}
\begin{figure}[!t]
    \centering
    \includegraphics[width=1.0\linewidth]{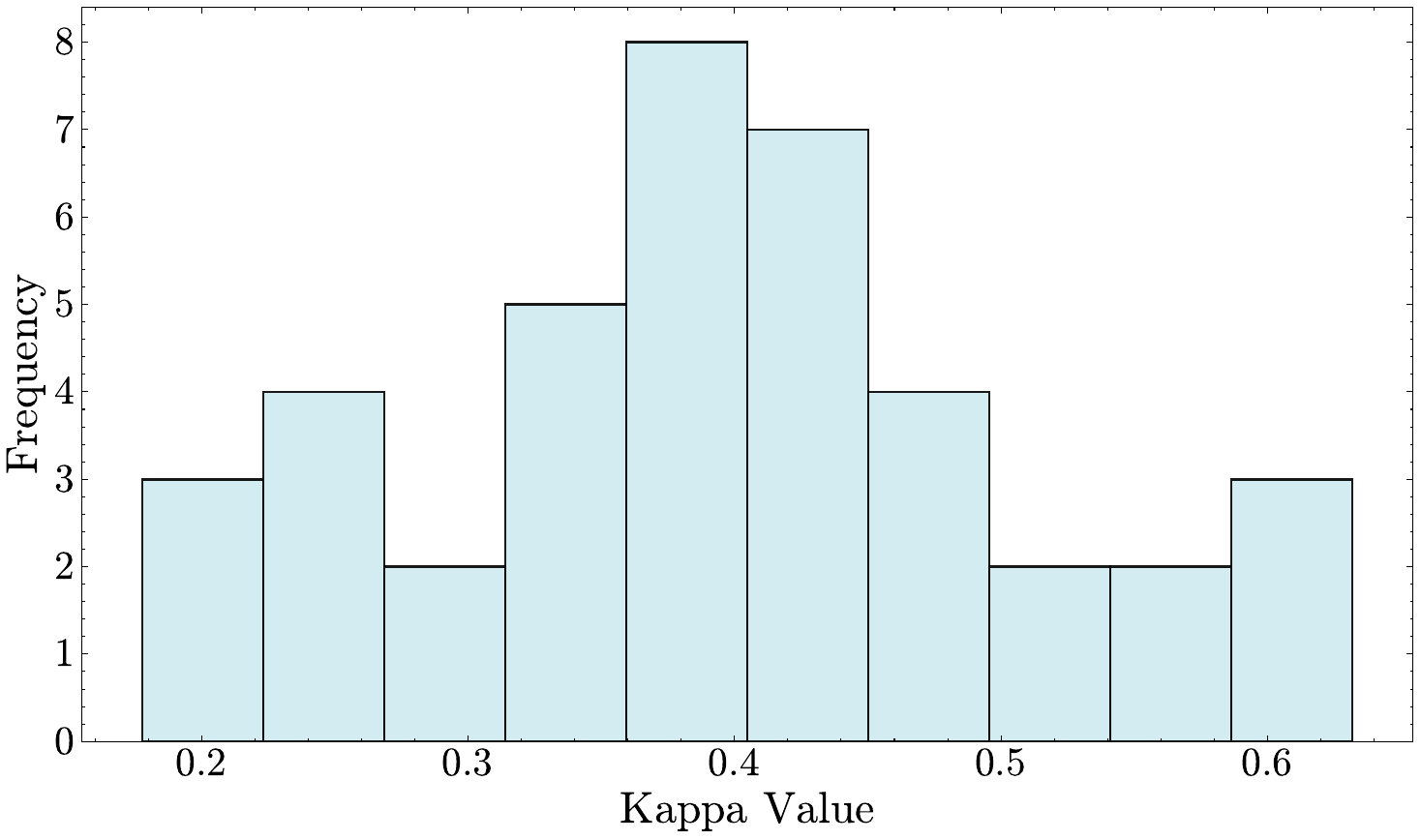}
    \caption{Kappa Scores distribution on TravelDest}
    \label{fig:kappa}
\end{figure}

We present the distribution of Cohen’s $\kappa$ scores for a subset of 42 queries from the \texttt{TravelDest} dataset (see Fig \ref{fig:kappa}). To assess the quality of the LLM-generated labels, we recruited domain experts in travel to manually annotate ground truth relevance for these queries. Comparison between the LLM-generated labels and expert annotations yields a Cohen’s $\kappa$ of 0.39, indicating \emph{fair agreement}.

\section{Examples}
In this section, we provide additional examples of various QR methods and results as shown in Tables~\autoref{tab:query_example1},\autoref{tab:query_example2},\autoref{tab:query_example3},\autoref{tab:query_example4}, and \autoref{tab:query_example5}.

\clearpage

\begin{table*}[h!]
\centering
\resizebox{\linewidth}{!}{
\small % Adjust font size
\begin{tabular}{|p{2cm}|p{12cm}|}
\hline
\textbf{Method} & \textbf{Reformulation / Results} \\
\hline
Q2E & extreme sports; hiking trails; rock climbing; water sports; skydiving\\
\textbf{Top Results} & \textit{Galapagos Islands, Queenstown (New Zealand), Cape Town, Denver, Hong Kong, \dots}\\
\hline
Q2D & \underline{Queenstown, New Zealand}: Known as the 'Adventure Capital of the World,' Queenstown offers a wide variety of adrenaline-filled activities set against stunning alpine scenery. Highlights include bungee jumping, skydiving, white-water rafting, mountain biking, and hiking. In winter, the region becomes a hub for skiing and snowboarding at nearby resorts. With its mix of extreme sports, outdoor adventures, and a lively town center, Queenstown is a premier destination for thrill-seekers of all kinds.\\ & \textit{Queenstown (New Zealand), New York City, Auckland, Santiago, Dunedin, \dots} \\
\hline
\textbf{EQR} & \underline{Mountain Adventures} - Cities that offer hiking, climbing, and skiing opportunities in nearby mountain ranges such as Queenstown (New Zealand) and Aspen. \underline{Water Sports} - Coastal cities known for exceptional surfing, diving, and sailing activities such as Honolulu and Gold Coast. Jungle Expeditions - Locations that provide guided tours and expeditions into dense jungles and rainforests such as Manaus and Belize City. \underline{Desert Safaris} - Cities that offer desert experiences, including dune bashing and camel rides such as Dubai and Abu Dhabi. Extreme Sports - Cities that host a range of extreme sports from bungee jumping to paragliding such as Queenstown (New Zealand) and Interlaken. \\
\textbf{Top Results} & \textit{Queenstown (New Zealand), Puerto Vallarta, Tenerife, Santiago,
Gold Coast, \dots} \\
\hline
\end{tabular}}
\caption{Query: Top cities for adventure seekers}
\label{tab:query_example1}
\end{table*}

\begin{table*}[h!]
\centering
\resizebox{\linewidth}{!}{
\small % Adjust font size
\begin{tabular}{|p{2cm}|p{12cm}|}
\hline
\textbf{Method} & \textbf{Reformulation / Results} \\
\hline
Q2E & youth-friendly activities; budget accommodations; group tours; adventure parks; cultural experiences\\

\textbf{Top Results} & \textit{Aarhus,
        San Francisco,
        San Diego,
        Washington, D.C.,
        Rochester (New York), \dots}\\
\hline
Q2D & \underline{New York City, USA}: As one of the world’s most iconic destinations, New York City offers a dynamic setting for high school graduation trips, combining celebration with cultural exploration. Landmarks such as Times Square, Central Park, and the Statue of Liberty provide memorable backdrops for group activities and photos. Students can enjoy Broadway shows, visit renowned museums like the American Museum of Natural History and the Met, and experience the energy of the city’s diverse neighborhoods. NYC’s mix of entertainment, history, and educational attractions makes it a popular choice for marking the milestone of graduation with both fun and enrichment.\\
\textbf{Top Results} & \textit{New York City, London, Manchester, Philadelphia, Rome, \dots
} \\
\hline
\textbf{EQR} & \underline{Adventure Activities}: Cities that offer exciting outdoor activities and adventures ideal for energetic young adults, such as Queenstown and Interlaken. \underline{Cultural Hotspots}: Cities rich in cultural experiences, museums, and historical sites, providing educational value, such as Rome and Athens. \underline{Beach Destinations}: Popular coastal cities with vibrant beach scenes and nightlife suitable for young travelers, such as Miami and Cancun.\\
\textbf{Top Results} & \textit{San Francisco, Milan, Shanghai, New York City, Athens, \dots
} \\
\hline
\end{tabular}}
\caption{Query: Cities for a high school graduation trip}
\label{tab:query_example2}
\end{table*}

\begin{table*}[h]
\centering
\resizebox{\linewidth}{!}{
\small % Adjust font size
\begin{tabular}{|p{2cm}|p{12cm}|}
\hline
\textbf{Method} & \textbf{Reformulation / Results} \\
\hline
Q2E & wellness centers; yoga retreats; meditation centers; hermal baths; sunshine beaches; nature reserve; peaceful countryside \\
\textbf{Top Results} & \textit{Palm Springs,
        Aruba,
        Cape Town,
        Gold Coast, \dots} \\
\hline
Q2D & \underline{Sedona, Arizona, USA}: Renowned for its striking red rock landscapes and vibrant arts community, Sedona is a premier destination for those seeking a rejuvenating retreat. The area is home to numerous wellness resorts, spas, and healing centers offering services such as massages, yoga, meditation, and holistic therapies. Visitors can also explore scenic hiking trails, vortex sites believed to promote spiritual renewal, and tranquil spots for reflection and relaxation. Sedona’s combination of natural beauty, peaceful atmosphere, and wellness-focused offerings makes it an ideal escape for rest and rejuvenation.\\
\textbf{Top Results} & \textit{Santo Domingo,
        São Paulo,
        Tenerife,
        Jakarta,
        Cabo San Lucas, \dots
} \\
\hline
\textbf{EQR} & 
\underline{Spa Retreats} - Cities known for offering luxurious spa services that combine relaxation with a variety of wellness treatments, perfect for rejuvenating the mind and body, such as Budapest and Bali. \underline{Nature Escapes} - Destinations surrounded by stunning natural landscapes, ideal for outdoor activities like hiking and sightseeing, and offering a peaceful break from the fast pace of life, such as Asheville and Queenstown (New Zealand). \underline{Beachfront Relaxation} - Cities with serene and picturesque beaches, perfect for enjoying sunbathing, swimming, and rejuvenating by the sea, such as Maldives and Honolulu.\\
\textbf{Top Results} & \textit{Mombasa,
        Santo Domingo,
        Aruba,
        Puerto Vallarta,
        Maldives, \dots} \\
\hline
\end{tabular}}
\caption{Query: Cities for a rejuvenating retreat}
\label{tab:query_example3}
\end{table*}

\begin{table*}[h]
\centering
\resizebox{\linewidth}{!}{
\small % Adjust font size
\begin{tabular}{|p{2cm}|p{12cm}|}
\hline
\textbf{Method} & \textbf{Reformulation / Results} \\
\hline
Q2E & quaint villages; cobblestone streets; local markets; artisan shops; scenic views; historic downtown; peaceful retreats; cultural festivals; bed and breakfasts; picturesque landscapes \\
\textbf{Top Results} & \textit{Albuquerque,
        Aurangabad,
        Aarhus,
        Ottawa,
        George Town (Malaysia), \dots} \\
\hline
Q2D & \underline{Bruges, Belgium}: Renowned for its beautifully preserved medieval architecture, winding cobbled streets, and picturesque canals, Bruges embodies the charm of a classic European small town. Visitors can explore historic landmarks such as the Belfry of Bruges and Basilica of the Holy Blood, enjoy leisurely boat rides along its scenic waterways, and stroll through quaint squares lined with cozy cafes and artisan chocolate shops. The city also boasts a rich artistic heritage, with museums showcasing Flemish masterpieces. Bruges’ intimate scale, storybook scenery, and welcoming atmosphere make it an ideal destination for travelers seeking a peaceful yet culturally rich escape.\\
\textbf{Top Results} & \textit{Amsterdam,
        Lisbon,
        Brussels,
        Tallinn,
        Aarhus, \dots
} \\
\hline
\textbf{EQR} & \underline{Historic Charm}: Towns that provide a rich sense of history, featuring well-preserved architecture and deep-rooted local traditions, perfect for cultural exploration, such as Bathurst (New Brunswick) and Ljubljana. \underline{Natural Beauty}: Small towns nestled in breathtaking natural surroundings, offering opportunities for outdoor activities like hiking, photography, and nature walks, such as Aspen and Queenstown (New Zealand). \underline{Cultural Festivals}: Towns renowned for their distinctive local festivals, giving visitors an authentic insight into regional culture and traditions, such as Edinburgh and Pamplona.
\\

\textbf{Top Results} & \textit{Riga,
        Aarhus,
        Albuquerque,
        Edmonton,
        Montevideo, \dots
} \\
\hline
\end{tabular}}
\caption{Query: Charming small town cities}
\label{tab:query_example4}
\end{table*}

\begin{table*}[h]
\centering
\resizebox{\linewidth}{!}{
\small % Adjust font size
\begin{tabular}{|p{2cm}|p{12cm}|}
\hline
\textbf{Method} & \textbf{Reformulation / Results} \\
\hline
Q2E & off-the-beaten-path; secluded; quiet towns; remote; less touristy; undiscovered; peaceful; small towns; hidden gems; tranquil \\
\textbf{Top Results} & \textit{São Paulo,
        Manchester,
        Brussels,
        Ibiza,
        Nice, \dots} \\
\hline
Q2D & \underline{Ljubljana, Slovenia}: Often overlooked in favor of larger European capitals, Ljubljana is a hidden gem that offers a peaceful and unhurried atmosphere ideal for travelers seeking a quieter experience. The city is renowned for its abundant green spaces, including Tivoli Park and the scenic Ljubljanica River, as well as its charming, pedestrian-friendly old town filled with pastel-colored buildings and riverside cafes. Far from the crowds of major tourist hubs, Ljubljana combines small-town intimacy with cultural richness, featuring medieval castles, open-air markets, and local art galleries. Its laid-back vibe, clean streets, and tranquil public spaces make it a perfect destination for those looking to explore Europe off the beaten path while enjoying a relaxed and authentic setting. \\
\textbf{Top Results} & \textit{Reykjavík,
        Helsinki,
        Ljubljana,
        Aarhus,
        Tallinn, \dots
} \\
\hline
\textbf{EQR} & \underline{Remote Locations}: Cities that are off the beaten tourist path, providing a sense of solitude and offering distinctive, memorable experiences, such as Iqaluit and Ålesund. \underline{Small Town Charm}: Smaller cities known for their peaceful streets, intimate atmosphere, and lack of large tourist crowds, making them ideal for a slower-paced getaway, such as Bathurst (New Brunswick) and Lethbridge. \underline{Nature Escapes}: Cities situated near expansive nature reserves and national parks, where visitors can easily disconnect from urban life and immerse themselves in the tranquility of the outdoors, such as Whitehorse and Aspen.
\\
\textbf{Top Results} & \textit{
        Brussels,
        Reykjavík,
        Ljubljana,
        Budapest,
        Venice, \dots
} \\
\hline
\end{tabular}}
\caption{Query: Best cities to avoid crowds}
\label{tab:query_example5}
\end{table*}

\end{document}